# Ultrafast Dynamics of Plasmon-Exciton Interaction of Ag Nanowire-Graphene Hybrids for Surface Catalytic Reactions

Qianqian Ding,[1, 2, 3, +] Ying Shi,[4,+] Maodu Chen,[3,+] Hui Li,[4,5,+] Xianzhong Yang,[2] Yingqi Qu,[6] Wenjie Liang,[2] Mengtao Sun[1,2,5,*]

1. Department of Applied Physics, School of Mathematics and Physics, University of Science and Technology Beijing, Beijing 100083, People's Republic of China
2. Beijing National Laboratory for Condensed Matter Physics, Beijing Key Laboratory for Nanomaterials and Nanodevices, Institute of Physics, Chinese Academy of Sciences, Beijing 100190, People's Republic of China
3. Key Laboratory of Materials Modification by Laser, Electron, and Ion Beams (Ministry of Education), School of Physics and Optoelectronic Technology, Dalian University of Technology, Dalian 116024, People's Republic of China
4. Institute of Atomic and Molecular Physics, Jilin University, Changchun 130012, People's Republic of China
5. Department of Physics, Liaoning University, Shenyang 110036, Liaoning, People's Republic of China
6. School of Physics, Peking University, Beijing 100871, People's Republic of China
* Corresponding Author. E-mail: mtsun@iphy.ac.cn (M. T. Sun).
[+] Contributed equally

Using the ultrafast pump-probe transient absorption spectroscopy, the femtosecond-resolved plasmon-exciton interaction of graphene-Ag nanowire hybrids is experimentally investigated, in the VIS-NIR region. The plasmonic lifetime of Ag nanowire is about 150 ± 7 femtosecond (fs). For a single layer of graphene, the fast dynamic process at 275 ± 77 fs is due to the excitation of graphene excitons, and the slow process at 1.4 ± 0.3 picosecond (ps) is due to the plasmonic hot electron interaction with phonons of graphene. For the graphene-Ag nanowire hybrids, the time scale of the plasmon-induced hot electron transferring to graphene is 534 ± 108 fs, and the metal plasmon enhanced graphene plasmon is about 3.2 ± 0.8 ps in the VIS region. The graphene-Ag nanowire hybrids can be used for plasmon-driven chemical reactions. This graphene-mediated surface-enhanced Raman scattering substrate significantly increases the probability and efficiency of surface catalytic reactions co-driven by graphene-Ag nanowire hybridization, in comparison with reactions individually driven by monolayer graphene or single Ag nanowire. This implies that the graphene-Ag nanowire hybrids can not only lead to a significant accumulation of high-density hot electrons, but also significantly increase the plasmon-to-electron conversion efficiency, due to strong plasmon-exciton coupling.

Surface plasmon resonance (SPR) is the resonant collective oscillation of conduction electrons at the interface between materials with negative and positive permittivities, which is stimulated by incident light[1]. SPR in sub-wavelength-scale nanostructures can be polaritonic or plasmonic[2]. After the plasmons are excited, in the first 1–100 femtoseconds (fs) following Landau damping, the thermal distribution of electron–hole pairs decays either through the re-emission of photons or through carrier multiplication caused by electron–electron interactions[3, 4]. During this short time interval $\tau_{nth}$, the hot carrier distribution is highly non-thermal. The hot carriers then redistribute their energy through electron–electron scattering processes on a timescale $\tau_{el}$ ranging from 100 fs to 1 ps[3, 4]. Since 2011, hot electrons generated from plasmon decay (we call it plasmon-induced hot electrons afterward) have been successfully applied to study plasmon-driven surface catalytic reactions monitored by the surface-enhanced Raman scattering (SERS) spectroscopy[5, 6] and the tip-enhanced Raman spectroscopy (TERS)[7]. Furthermore, plasmon-induced hot electrons have been found to play a key role in plasmon-driven reduction reactions[8]. The surface-plasmon-to-hot-electron conversion efficiency is typically smaller than 1%[9]. Therefore, it is a great challenge to increase the probability and efficiency of plasmon-driven chemical reaction via increasing the efficiency of the plasmon-to-electron conversion. Recently Ag Nanoparticle-TiO$_2$ hybrid had been used to study the plasmon-driven chemical reaction.[10, 11] Furube and coworkers[12, 13] experimentally revealed the mechanism of ultrafast plasmon-induced electron transfer from Gold nanodots into TiO$_2$ nanoparticles, by using femtosecond transient absorption spectroscopy with an IR probe.[12] They directly observed plasmon-induced electron transfer from 10 nm Gold nanodots to TiO$_2$ nanoparticles, and the reaction time was within 240 fs.[12]

Graphene, a single atomic layer of graphite, has attracted vast interest owing to its unique properties since its discovery in 2004[14]. It has been used as a substrate for graphene-enhanced Raman scattering (GERS) spectroscopy[15]. Recently, two vibrant and rich fields of investigation regarding graphene and plasmonics have been combined[16]. Graphene-mediated SERS (G-SERS) substrates[15, 17-24], have been successfully applied in plasmon-graphene co-driven chemical reactions[25-27]. The surface-plasmon-to-hot-electron conversion efficiency revealed by femtosecond transient spectroscopy increases by 8.7 times in reduced graphene-coated gold nanoparticles[28]. However, to the best of our knowledge, the mechanism of plasmon-graphene co-driven chemical reactions is not clearly revealed yet. For examples, how to reveal the ultrafast plasmon-exciton interaction between metal and graphene? How to monitor and reveal the ultrafast dynamics process of the plasmon-induced hot electrons transferring to graphene? What is the time scale of it? Why can the efficiency of the catalytic reaction be significantly increased by the hybrids system? All of these questions are urgent to be answered in this field, which should be explicitly pointed out. We tried to answer part of them in our manuscript.

In this paper, novel optical graphene-Ag nanowire hybrid materials are fabricated, and the advantages of this hybrid device are shown through ultrafast pump-probe transient absorption spectroscopy and surface catalytic reactions on this hybrid nanostructure. The experimental results indicate that graphene can strongly harvest hot electrons generated from Ag plasmon decay, which can not only lead to a significant accumulation of high density hot electrons, but also prolong the lifetime of these hot electrons from femtosecond to picosecond, through Ag plasmon-enhanced graphene plasmon. Furthermore, the surface-plasmon-to-hot-electron conversion efficiency is significantly enhanced. Our results show that the graphene-Ag nanowire hybrids with plasmon-exciton interaction can act as a good candidate in spectral censoring, catalysis, and potential applications in nano-photonic devices, energy and environments.

### Results

#### Femtosecond-resolved plasmonic dynamics of Ag nanowire

Fig. 1(a) is the optical image of as-prepared Ag nanowire and 1(b) is the SEM images of the Ag nanowire, which are about 200 nm in diameter. We first measured the lifetime of plasmon for Ag nanowire using the ultrafast pump-probe transient absorption spectroscopy. Generally, transient signals originate from three parts, i.e., ground state bleaching signal, stimulated emission and absorbance of excited states. Fig. 1(c) demonstrates the energy distribution of collective oscillation of surface electrons when the surface plasmon is excited by the 400 nm pump laser on the quartz substrate; furthermore, the fitted curve in Fig. 1(d) reveals that the lifetime of plasmon in the Ag nanowires is $150 \pm 7$ fs. Besides, the $\Delta A$ in Fig. 1(d) represents the difference of absorbance spectra[29]. The details of the fitted lifetimes of the plasmons in Ag nanowires are shown in Table 1. Note that, we get the corresponding fitted lifetimes according to the fitted function[30], see Table S1.

#### Femtosecond-resolved dynamics of single layer graphene

The SEM image of the graphene on $SiO_2$/Si substrate is shown in Fig. 2(a), and the Raman spectrum in Fig. 2(b) qualitatively shows that it is a single layer of graphene, from the ratio of Raman intensity ($I_{2D}/I_G \approx 2.23$)[31]. The lifetime of graphene plasmon is measured by using the ultrafast pump-probe transient absorption spectroscopy at VIS region (Fig. 2(c)). Note that the substrate is quartz during the measurement of transient absorption spectroscopy. The fitting curve in Fig. 2(d) reveals the lifetime of graphene plasmon is fitted for 532 nm wavelength. There are two processes, i.e., $275 \pm 77$ fs and $1.4 \pm 0.3$ ps, respectively (see Table 1). The fast process at fs scale is for the excitation of electrons in graphene. And the slow process at ps scale corresponds to the excited electrons interacting with phonons in graphene. The ultrafast pump-probe transient absorption spectroscopy for graphene at NIR region is plotted in Fig. 2(e). The fitting curve in Fig. 2(f) exhibits the lifetime of graphene plasmon is fitted for 1103 nm wavelength. There are two processes, $320 \pm 46$ fs and $2.5 \pm 0.6$ ps, respectively (see Table 1). The fast process at fs scale is for the excitation of electrons in graphene, and the slow process at ps scale is for the excited electrons interacting with phonons in graphene. Comparing the two processes at VIS and NIR, it is found that with the increase of the laser wavelength, the time scales for the two processes are enlarged, respectively. This phenomenon implies that the lower the energy of the laser, the slower the process of these two dynamic processes.

#### Femtosecond-resolved dynamics of plasmon-exciton coupling of graphene-Ag nanowire hybrids

The SEM image of a single Ag nanowire coated by monolayer graphene can be seen in Figure 3(a). We measure the time of the charge transfer between Ag nanowire and single layer graphene on the quartz substrate in the VIS region, see Fig. 3(b), and there are two dynamic processes with different time scales. The fitted time of hot electrons transferring from Ag nanowire to graphene is about 534 ± 108 fs. It is slower than the reaction time within 240 fs for the plasmon induced electron transfer from 10 nm gold nanodots to $TiO_2$ nanoparticles in IR region.[12] In this time scale about 534 ± 108 fs, the hot carriers redistribute their energy through electron–electron scattering, which is confirmed by Figure 3(b). Figure 3(b) demonstrates the energy of hot electrons distributed along the whole measured range of wavelength, which reveals that some plasmon-induced hot electrons lost their kinetic energy in the process of transferring to graphene. The fitted curve in Figure 3(c) also indicates that the time of plasmon-induced hot electrons interacting with phonons in graphene is about 3.2 ± 0.8 ps. So, for the hybridized plasmonic system of graphene-Ag nanowire, the lifetime of plasmon-induced hot electrons in graphene is about 3.2 ± 0.8 ps, which is significantly longer than that of isolated Ag nanowire, since the period of collective electron oscillation (CEO) for isolated Ag nanowire is within 150 fs.

We also measured the time of charge transfer between Ag nanowire and single layer graphene in the NIR region, as shown in Fig. 3(d), which demonstrates that the kinetic energy of hot electrons can be down to about 0.8265 eV (1500 nm). In fact, the energy of hot electrons is equal to the laser energy if there is no energy loss. The energy distribution of hot electrons is the Fermi-Dirac distribution, which is strongly dependent on the intensity of the laser. Due to the non-conservation of momentum in the system[32, 33], the hot electrons have high energy and occupy the entire region of $E_F < \varepsilon < E_F + h\nu$ in the system, where $E_F$ is the Fermi energy level.

There are two dynamic processes with different time scales in the NIR region. The fitted time of hot electrons transferring from Ag nanowire to graphene is about 780 ± 92 fs. It is found that the reaction time of hot electron from Ag nanowire to graphene in the NIR region is slower that in the VIS region, due to lower kinetic energy in the NIR region. The fitted curve in Figure 3(e) also reveals that the time of plasmon-induced hot electrons on graphene interacting with phonons in graphene is about 3.9 ± 0.9 ps (shown in Table 1), which is slower than that in VIS region, due to better optical absorption for graphene in the NIR region. Therefore, for the graphene-Ag nanowire hybridized plasmonic system, the lifetime of plasmon-induced hot electrons on graphene is about 3.9 ± 0.9 ps in the NIR region.

**Surface catalytic reaction induced by plasmon**

One of the advantages of graphene-Ag nanowire hybrids is that, graphene plasmon enhanced by metal plasmon can be used to drive chemical reactions on this hybrid nanostructure The surface catalytic reaction of 4-nitrobenzenethiol (4NBT) to p,p'-dimercaptoazobenzene (DMAB) without graphene has already been clearly investigated[34]. Thus, it is a best candidate for us to show the advantages of the optical materials consisting of graphene-Ag nanowire hybrids fabricated in the current work.

To highlight the advantages of plasmon-graphene hybrids for the catalytic reactions, we first measured the Raman spectra of 4NBT and DMAB powders, and SERS spectra of 4NBT at the junction of tip-to-tip Ag nanowires (SEM image shown in Fig. 1(b)) without single layer graphene, under the excitation light of 532 nm with laser power of 1.23 mW (10 %). Figure 4(a), 4(b) and 4(c) show the normal Raman spectrum of 4NBT powder, SERS spectrum of 4NBT and Raman spectrum of DMAB power, respectively. The Raman spectra from 1300 cm$^{-1}$ to 1500 cm$^{-1}$ in Fig. 4a-c are what we are interested in the catalytic reactions. It is found that the SERS spectrum of 4NBT (Fig. 4(b)) is different from Fig. 4(a) or Fig. 4(c), which reveals that a partial chemical reaction must have occurred with a strong Raman background, showing that the reactant, product and intermediate species coexisted during the reaction. This implies that single Ag nanowire itself is not the best candidate for the plasmon-driven reduction reactions.

### Surface catalytic reaction on monolayer graphene

Now we present the results of our investigation on the graphene-driven reduction reactions of 4NBTs to DMAB on the GERS substrate. The SEM image is shown in Fig. 2(a), and the Raman intensity of graphene at 2687 cm$^{-1}$ is larger than that at 1596 cm$^{-1}$, proving that the monolayer graphene is present. The GERS spectra are presented in Fig. 5(a), where Fig. 5(b) shows the enlarged view of the spectra in the fingerprint region of Fig. 5(a), which exhibits that the Raman intensities of molecules in 1000-1600 cm$^{-1}$ range are extremely weak. The partially enlarged sections in Fig. 5(b) elucidate the contribution of the laser power. The Raman intensity of N-O vibrations of 4NBT at 1325 cm$^{-1}$ gradually decreases and then disappears with increasing laser power, while the Raman intensities of DMAB at 1390 and 1432 cm$^{-1}$ gradually increase as the laser power increases, see the qualitatively analysis in Figure S1. These observations imply that graphene can drive the surface reduction reactions and the reaction dynamics can be controlled by tuning the laser intensity. The DMAB Raman peak at 1143 cm$^{-1}$ due to β(C-H) cannot be observed because of the strong background of graphene. We propose that photo-induced hot electrons generated from graphene provide the kinetic energy and the electrons for the reduction reactions from 4NBTs to DMAB. The density and efficiency of hot electrons provided by the graphene are strongly dependent on the laser power. However, the intensity of Raman spectra in the chemical reactions is weak, compared with the strong graphene Raman signal. More electrons with higher kinetic energy are needed to increase the reaction probability and efficiency of graphene-driven reduction reactions when excited with weak laser power.

### Catalytic reaction on graphene-Ag nanowire hybrid system

The SEM image of graphene-Ag nanowire hybrids is presented in Fig. 3(a), which shows that the Ag nanowires is approximately 200 nm in diameter, and is covered with monolayer graphene. Figure 6 presents the measured results of the laser intensity-dependent plasmon-graphene co-driven reduction reactions of 4NBT on the graphene-Ag nanowire composite. Even at a weak laser intensity (0.1%, 0.01 mW), the conversion of 4NBTs to DMAB can be nearly 100% completed by plasmon-graphene hybrids (because the characteristic peak of 4NBT at 1325 cm$^{-1}$ corresponding to the N-O$_2$ vibrational modes of 4NBT almost disappears completely), which exhibits that the plasmon-graphene hybrids can provide a high-density hot electrons, and thus that leads to a highly efficient catalytic reaction. It is clearly observed that the intensities of Raman peaks of graphene at 1595 and 2687 cm$^{-1}$ gradually increase with the increase of the laser power. When the laser power reaches 100% (11 mW), the Raman intensity at 2687 cm$^{-1}$ becomes larger than that at 1595 cm$^{-1}$, $I_{2D}/I_G \approx 1.13$. Thus, we conclude that 1) the catalytic reaction is already highly efficient at weak laser power, and the reaction probability can reach 100% by increasing the laser power; 2) the intensities of the graphene Raman peaks at 1595 and 2687 cm$^{-1}$ also increase gradually, and for sufficiently high laser intensities, the Raman intensity at 2687 cm$^{-1}$ become larger than that at 1595 cm$^{-1}$, which demonstrates that the catalytic reaction indeed occurs on the monolayer graphene-coated Ag nanowires. Note that Raman intensities of DMAB are weaker compared with the Raman intensities of graphene, the reason is that some of the DMAB may be desorbed from the surface of graphene when radiated at enough strong laser. There are two superposed Raman peaks between graphene and DMAB around 1595 cm$^{-1}$, which leads to $I_{2D}/I_G > 1$, but is less than 2.3, when the laser power is 100%.

**Discussion**

Novel graphene-Ag nanowire hybrid optical materials are fabricated and probed using the ultrafast pump-probe transient absorption spectroscopy, and then exploited as the G-SERS substrates to investigate the typical surface reductions of 4NBTs to DMAB. Firstly, single-layer graphene can significantly improve the surface-plasmon-to-hot-electron conversion efficiency. Secondly, with the assistance of graphene, the lifetime of hot electrons is greatly increased from femtosecond to picosecond. As a consequence, this G-SERS substrate significantly increases the probability and efficiency of the surface catalytic reactions of 4NBTs to DMAB co-driven by graphene-Ag nanowire hybridization, compared with the same kinds of reactions but driven individually either by graphene or single Ag nanowires. The graphene-Ag nanowire hybrids, with higher conversion efficiency and longer lifetime of plasmon-induced hot electrons, can be used as a source of hot electrons for catalysis and might be potentially applied as an analytical tool for observing such plasmon-driven reactions.

**Method**

Large-area monolayer graphene was synthesized on copper foils (Alfa Aesar, 25 μm thick, 99.999% purity) using cold-wall low-pressure chemical vapour deposition (CVD) [36]. The copper foil was first soaked in acetic acid for 5 min to remove copper oxide and then annealed in H$_2$ gas at 1000 °C for 5 min. In the initial stage of the reaction, the reaction gaseous mixture Ar: H$_2$: CH$_4$ with a flow rate of 960: 40: 20 sccm was introduced into the tube. The reaction took place for 5 min at 1000 °C. Finally, the CVD tube was cooled to 150 °C under an Ar: H$_2$ environment.

The silver nanowires were prepared through chemical fabrication[37]. Under continuous magnetic stirring, 167 mg poly (vinyl pyrrolidone) (PVP) was dissolved in 15 mL of ethylene glycol (EG) solution. Then, 0.2 mL of the silver nitrate solution (1.5 M) was added. The mixture solution was kept at 70 °C for 30 min in an oil bath and then kept at 150 °C for 90 min. The final product was diluted with an ethanol solution and centrifuged five times to remove EG and PVP. The diluted Ag nanowires were dropped on the marked SiO$_2$/Si substrate and then dried in the air at room temperature.

Polymethyl methacrylate (PMMA) was used to help transfer the graphene from the copper foil to Ag nanowires on the SiO$_2$/Si substrate. First, a PMMA layer was spin-coated (3000 rpm, 1 min) on the graphene-covered copper foil. The copper foil was then etched away by ferric chloride solution (0.5 M) for 4 h. The PMMA film with graphene was soaked in deionized water four times and then transferred from the water to the substrate with Ag nanowires. Finally, the PMMA was dissolved in an acetone solution and then the sample was washed in an ethanol and deionized water.

The G-SERS substrates were immersed in a 5×10$^{-5}$ M 4-NBT ethanol solution for more than 12 h. The substrates were then washed with ethanol for 5 min, and dried with N$_2$ gas. All SERS spectra were measured by using a Renishaw inVia confocal Raman spectrometer with a 532 nm laser (11 mW, 100 % power) as the excitation light. Each Raman spectrum was recorded with an accumulation time of 10 s.

The morphologies of the substrate system consisting a single-layer graphene-Ag nanowire were characterized by using a scanning electron microscope (SEM, Hitachi S-4800). The setup of the transient absorption spectroscopy has been reported in the previous paper[38]. The femtosecond laser used was the Libra (Coherent) an all-in-one ultrafast oscillator and a regenerative amplifier laser system. The Coherent Vitesse served as the seed laser which produced 4 mJ pulses at a repetition rate of 1000 Hz. The output wavelength of the system was 800 nm where the full width at half maximum (FWHM) was 50 fs. The fundamental beam of a Ti: sapphire laser was separated into two beams in the ratio 9: 1. The more intense beam was used for generating the second harmonic ($\lambda_{ex}$=400 nm) of the fundamental laser by a 0.5 mm BBO (b-BaB$_2$O$_4$, Fujian Castech Crystals Inc. China). The 400 nm beam was attenuated to a power of approximately 3 μJ per pump pulse at the sample. The excited states were probed by using the other beam generated to pass through a controlled delay line (ILS250CCL, France) and focus onto a sapphire plate to generate a sub-picosecond white-light super-continuum. The pump and probe beams were incident on the sample at a small angle ($\theta \leq 5°$). Samples were placed in the beam path where the beam diameter was 300 μm. All of the above experimental measurements were performed at room temperature (295 K). The sample of Ag nanowire-graphene was placed on the quartz substrate for the transient absorption spectroscopy.

### ACKNOWLEDGEMENTS

This work was supported by the National Nature Science Foundation of China (grant nos. 91436102, 11374353, 11574115, 11374045, 21471039, 11374342, 11474141 and 11544015), National Basic Research Program of China (Grant number 2016YFA02008000), the Program for New Century Excellent Talents in University (grant no. NCET-12-0077), and the Program of Liaoning Key Laboratory of Semiconductor Light Emitting and Photocatalytic Materials. M. Sun thanks for the fanatical support by University of Science and Technology Beijing.

### Author contributions

M. Sun conceived and designed the experiments. Q. Ding, Y. Shi, H, Li and M. Sun performed the experiments. M. Sun, Q. Ding, Y. Shi, H. Li and M. Chen analysed the data. Q. Ding, Y. Shi and H. Li contributed materials/analysis tools. M. Sun wrote the paper. Y. Qu, X. Yang and W. Liang involved discussion and revised manuscript. All authors reviewed the manuscript.

### Competing financial interests

The authors declare no competing financial interests.

| System | Spectrum range of probe light | Lifetime of fast process ($\tau_1$ ($fs$)) | Lifetime of slow process ($\tau_2$ ($ps$)) |
|---|---|---|---|
| Ag nanowire | Vis | 150 ± 7 (4.4 %) | |
| Graphene | Vis | 275 ± 77 (22.4 %) | 1.4 ± 0.3 (27.9 %) |
| Graphene | NIR | 320 ± 46 (14.5 %) | 2.5 ± 0.6 (22.8 %) |
| AgNW/Graphene | Vis | 534 ± 108 (20.2 %) | 3.2 ± 0.8 (26.1 %) |
| AgNW/Graphene | NIR | 780 ± 92 (11.8 %) | 3.9 ± 0.9 (24.3 %) |

**Table1 | The fitted lifetimes of Ag nanowire, monolayer graphene and the Ag nanowire-graphene hybrid detected in the visible region, respectively; monolayer graphene and the Ag nanowire-graphene hybrid detected in the near-infrared region, respectively. The proportion is the deviation percentage, which ensures the accuracy of fitted parameters.**

**Figure Captions**

**Figure 1 | SEM imaging of Ag nanowires and ultrafast pump-probe transient absorption spectroscopy**. **(a)** The bright field optical image of Ag nanowires. **(b)** The high-resolution SEM image of Ag nanowires. **(c)** Ultrafast pump-probe transient absorption spectroscopy of Ag nanowires excited by 400 nm laser. **(d)** The fitted dynamic curve at 532 nm.

**Figure 2 | SEM imaging of single layer graphene and ultrafast pump-probe transient absorption spectroscopy.** **(a)** SEM imaging of graphene on $SiO_2$/Si substrate. **(b)** Raman spectrum of graphene in Fig. 1(a). **(c)** Ultrafast pump-probe transient absorption spectroscopy (VIS) of graphene excited by 400 nm laser. **(d)** The fitted dynamic curve at 532 nm. **(e)** Ultrafast pump-probe transient absorption spectroscopy (NIR) of graphene excited by 400 nm laser. **(f)** The fitted dynamic curve at 1103 nm.

**Figure 3 | SEM imaging of a single Ag nanowire veiled with monolayer graphene and ultrafast pump-probe transient absorption spectroscopy.** **(a)** SEM image of a single Ag nanowire coated by monolayer graphene. **(b)** Ultrafast pump-probe transient absorption spectroscopy of hybrid graphene-Ag nanowire excited by 400 nm laser. **(c)** The fitted dynamic curve at 532 nm. **(d)** Ultrafast pump-probe transient absorption spectroscopy of hybrid graphene-Ag nanowire in NIR region. **(e)** The fitted dynamic curve at 1103 nm.

**Figure 4 | Surface catalytic reactions on the tip-to-tip Ag nanowire. (a)** Raman spectrum of 4NBT powder. **(b)** SERS spectrum of 4NBT adsorbed on single Ag nanowire. **(c)** Raman spectrum of DMAB powder. The laser wavelength is 532 nm with the intensity of 10 % (1.23 mW) during the measurement of SERS spectra.

**Figure 5 | Surface catalytic reactions on monolayer graphene. (a)** Laser power-dependent GERS spectra of 4NBT under an illumination of 532 nm radiation. **(b)** Enlarged spectra of the spectra from 1000 $cm^{-1}$ to 1650 $cm^{-1}$ in Fig. 5(a).

**Figure 6 | Laser power-dependent plasmon-graphene co-driven reductions of 4NBT on the graphene-Ag nanowire hybrid nanostructure, with an excitation source of 532 nm.**